\documentclass[twocolumn,twoside,slac_two]{revtex4}
\usepackage{graphicx}
\usepackage{fancyhdr}
\begin{document}

\title{ATLAS Great Lakes Tier-2 Computing and Muon Calibration Center Commissioning}

%

\author{Shawn McKee}
\affiliation{Department of Physics, University of Michigan, Ann Arbor, MI 48109, USA}

\begin{abstract}
Large-scale computing in ATLAS is based on a grid-linked system of tiered computing
centers. The ATLAS Great Lakes Tier-2 came online in September 2006 and now is
commissioning with full capacity to provide significant computing power
and services to the USATLAS community. Our Tier-2 Center also host the Michigan Muon Calibration Center which is responsible for daily calibrations of the ATLAS
Monitored Drift Tubes for ATLAS endcap muon system. During the first LHC beam period in 2008 and following ATLAS global cosmic ray data taking period, the Calibration Center received a large data stream from the muon detector to derive the drift tube timing offsets and time-to-space functions with a turn-around time of 24 hours. We will present the Calibration Center commissioning status and our plan for the first LHC beam collisions in 2009.
\end{abstract}

\maketitle

\thispagestyle{fancy}


\section{Introduction}
The LHC is scheduled to restart by the end of 2009, opening up an exciting period of intensive effort by physicists all over the world.  As a result there will be a large amount of data from the LHC detectors which will need to be carefully analyzed, requiring very significant storage and computational power to process.  Also, since this data comes from new detectors, much of the initial effort will be focused on understanding the detector responses and behavior in detail; calibrating, aligning and verifying each detector subsystem. In this paper I will describe the work undertaken at the University of Michigan to commission both our computing and ATLAS muon calibration centers.   

\section{ATLAS Great Lakes Tier-2 (AGLT2)}
\begin{figure}[h]
\centering
\includegraphics[width=80mm]{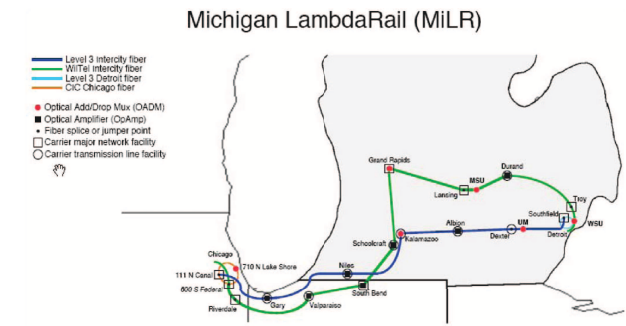}
\caption{Michigan LambdaRail(MiLR) is composed of two independent fiber paths, each capable of supporting multiple 10GE network connections and thereby redundantly interconnecting the AGLT2 sites with an international peering point at StarLight in Chicago.} \label{MiLR}
\end{figure}

The ATLAS Great Lakes Tier-2 Computing Center, AGLT2\cite{AGLT2}, is one of five ATLAS Tier-2 centers in the United States and is funded by the National Science Foundation to help meet the computing needs of the ATLAS experiment.  The center is physically distributed between the University of Michigan in Ann Arbor and Michigan State University in East Lansing.  Even though the site is physically split, the Tier-2 center appears logically as a single center with redundant 10 gigabit ethernet network connections to the world (see figure~\ref{MiLR}).  Both computer and storage nodes are transparently accessible, independent of their physical location.  

AGLT2 currently has over 1800 ``job slots'' available for running computational tasks.  A ``job slot'' conceptually represents a CPU core capable of running a program with its associated memory and access to input and output locations.  Typically, input files are staged to local storage on the node which runs the job, a calculation (filter, transform, etc) is  performed on the data and the output is sent to a specific destination (usually grid-aware storage).  

The grid storage system in use at AGLT2 is dCache/Chimera v1.9.4-3\cite{dCache}.   We currently have 524 TB of production storage running dCache and another 200 TB of storage in use for prototyping and testing future storage systems (Lustre, GlusterFS, etc.).  The storage areas in dCache utilize space reservations in the form of ``space-tokens'' which reserve and track storage space. 

The matchmaking between queued jobs and available job slots is handled by a job scheduler.  At AGLT2 we use Condor\cite{Condor} to manage job queueing, scheduling and related policy implementation.  With Condor we are able to share and prioritize our computational resources based upon our requirements.   This is especially important for the allocation of required resources for the ATLAS Muon Calibration Center discussed in section~\ref{MuonCal}.  

\subsection{AGLT2 Status and Performance}
The AGLT2 has been in operation since Fall of 2006, supporting both ATLAS production and user and group analysis activities.  The Tier-2 center has about two full-time equivalents (FTE) of manpower and involves contributions from six people for operating and maintaining the center.

In June 2009 the World-wide LHC Grid (WLCG) conducted an exercise (STEP09) designed to approximate conditions during the LHC restart at the end of the year.  Our Tier-2 performed very well, running the second most number of analysis jobs of any Tier-2 worldwide and moving the second most number of data files for Tier-2s worldwide.  We averaged over 625 megabytes/sec data transfer bandwidth for the eight day duration of STEP09.

Also, during 2009 our site has regularly been achieving 250-600\% of our WLCG  commitment for CPU-hours delivered. Typical monthly availability and reliability has been 95\% or above. Summarizing the Tier-2 status, AGLT2 has been operating very well and initial commissioning tests have verified the system is ready for LHC startup.   

However AGLT2 is unique amoung USATLAS Tier-2s, in that it has the additional responsibility of being an ATLAS Muon Calibration Center.

\vspace{-15pt}
\section{Michigan Muon Calibration Center}\label{MuonCal}

The University of Michigan is one of three ATLAS Muon Calibration Centers, (Max-Planck Institute/Ludwig Maximilians University and Rome I are the other two).  The muon calibration centers are intended to provide the needed calibration and alignment for the muon MDT (monitored drift tube) subsystem. Understanding the detector data can be complicated by changing conditions; gas composition, temperature, humidity, and voltage variations can impact the interpretation of the muon subsystem data.  The centers are designed to provide quick calibration and alignment so that the Tier-0 center at CERN can create the initial ESDs (event summary data) from the raw data.   The goal of the calibration centers is to continually provide this data within 24 hours, although the requirement is 48 hours because that is when the Tier-0 does the first pass reconstruction.

At Michigan we have implemented the calibration center as a logical subset of the existing AGLT2.  Because the storage services and computational requirements for the calibration center are very similar to that of the Tier-2 it is easy to accomodate the requirements of the calibration center using what already exists for the Tier-2 as a starting point.  Having the calibration center defined in this way significantly reduces the required manpower to manage and operate the center. Currently about 1.5 FTE is assigned to the calibration center but four people contribute to its operation and management.  There are additional unique requirements that the calibration center imposes which we must address.   That will be discussed in section~\ref{Unique} below.

The primary requirements for the center are high-priority access to at least 100 CPUs, a storage element and associated space for muon calibration data and software to manage and monitor the calibration tasks and insure their timely completion.

\subsection{Muon Calibration Overview}
The ATLAS muon system utilizes a number of technologies to accurately measure muons.  The MDT (monitored drift tubes) subsystem, which is the focus of the calibration centers, are composed of 1-5 meter, 2.54 cm diameter aluminum tubes which have a tungsten wire strung down their center and held by precision tube plugs at each end.  The tubes are filled with a pressurized Argon-C0$_2$ gas mixture at 3 atmospheres and the wire is held at 3090 V.   Muons passing through a drift-tube ionize the gas and the cluster of freed electrons drift toward the wire at the center of the tube.  The resulting avalanche as the electrons reach the wire creates a pulse which stops a TDC counter, providing a ``drift time''.

Calibration for the MDTs require that we determine the $T_0$ for each tube (the time corresponding to 0 drift time equivalent to the muon intersecting the wire) as well as the time-to-space function which maps drift time to a distance from the wire.

\begin{figure}[h]
\centering
\includegraphics[width=80mm]{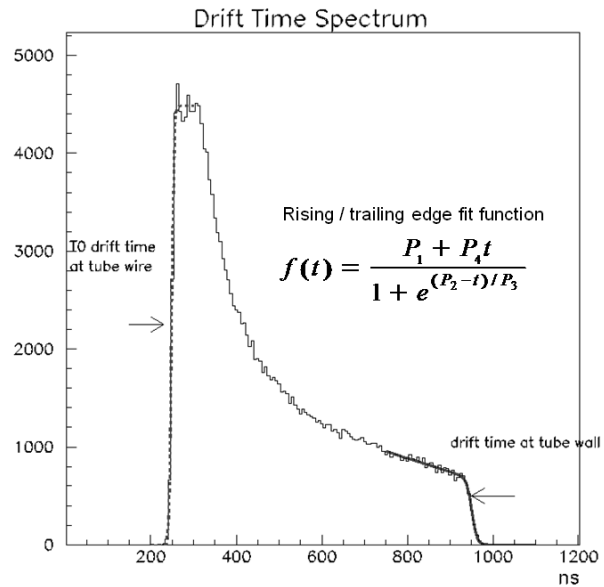}
\caption{A typical MDT drift time spectrum showing the definition of $T_0$ as well as $T_{max}$.} \label{DriftTime}
\end{figure}

\begin{figure}[h]
\centering
\includegraphics[width=80mm]{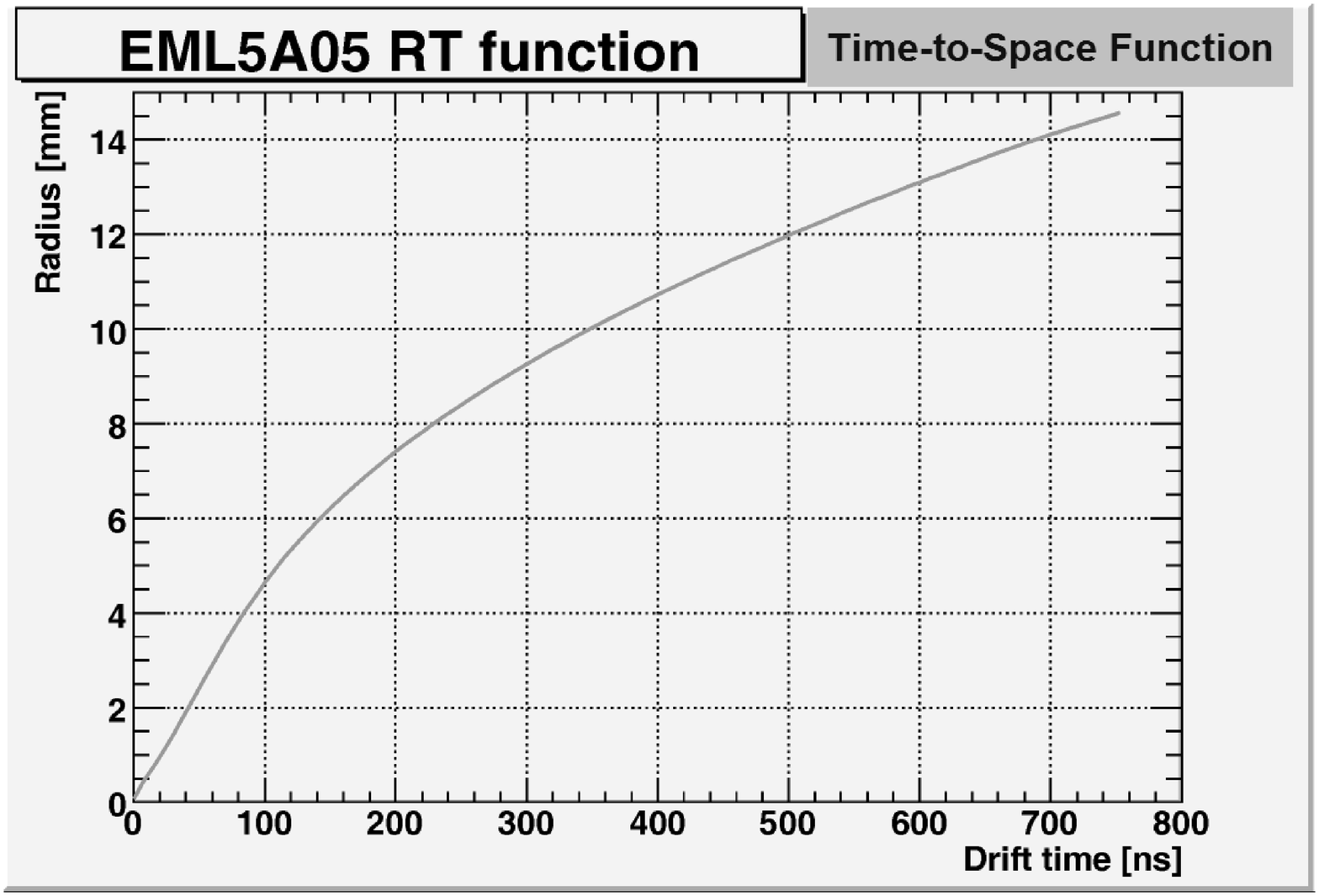}
\caption{A typical MDT ``time-to-space'' function maps drift times into radius of closest approach for the muons.} \label{TimeSpace}
\end{figure}

A typical MDT time spectrum is shown in figure~\ref{DriftTime}.   The $T_0$ values represent the combination of offsets due to electronic and cabling delays.  The time-to-space function is shown in figure~\ref{TimeSpace} and provides the mapping between the observed drift time and the radius of closest approach for the muon.  This function is very sensitive to the gas composition, the voltage on the wire, the local integrated B-field, temperature and humidity variations and the presence of contaminants in the gas.  Accurately determining the $T_0$ and time-to-space functions during ATLAS data-taking runs is the primary task for the muon calibration effort.

\vspace{-20pt}
\subsection{Calibration Data Flow}

The real-time aspect of data processing required for the calibration centers imposes additional requirements on the underlying computing infrastructure.  The data flow is critical for achieving timely calibration results.  All three calibration centers are configured to receive muon calibration data streams directly from the Tier-0 at CERN using the regular ATLAS DDM (Distributed Data Management) system DQ2\cite{DQ2}.

Approximately $10^4$ tracks per tube are required to accurately determine $T_0$ values for the MDTs.  Regular ATLAS trigger streams do not have enough muon tracks to allow us to calibrate the drift tubes in a timely manner.   In order to get adequate statistics a {\it muon calibration data stream} has been implemented.  ATLAS data is recorded by the Trigger/DAQ (Data AcQuisition) system which has a 3-level structure to select which collision data to store.  The system performs a sophisticated pattern recognition at level-2 to select only data with high-momentum muons for the muon calibration stream.   This data is sent to the three muon calibration centers and provides approximately 10 times as many muons as the ordinary data stream -- about $10^8$ per day (or 100 MB/day) at a luminosity of $10^{33}$ cm$^{-2}$ s$^{-1}$.

Because of the critical importance of receiving the calibration data stream in a timely manner and the trans-Atlantic location of the Michigan calibration center, we have implemented a secondary (backup) data path in case of problems with the primary distribution path.   This involves having an additional storage node located at CERN connected via a virtual circuit to our calibration center at Michigan.  This provides an alternate path and data source for failures in the primary distribution stream.

\subsection{Unique Requirements}\label{Unique}

While the calibration center requirements are significantly addressed by existing capabilities within the Tier-2 center there are a few unique requirements we must provide:
\begin{itemize}
\item Prioritized access to computing and storage to meet the real-time deadlines for calibration data
\item Network circuits to support data distribution to and from the center
\item A special {\bf CALIBDISK} Storage Resource Manager 2.2 complient storage area for incoming muon data
\item A local ORACLE server with STREAMS replication back to CERN's ORACLE server
\item Workflow management software to manage and track the calibration effort
\end{itemize}

To meet the priority requirements we have utilized Condor configuration options to allow incoming calibration jobs to have the highest priority access to job slots.  Given the large number of job slots at AGLT2, it only takes between 4-40 minutes to ramp up to a full set of 100 jobs (for 12 hour jobs the average waiting time for a slot is 24 seconds).  In our commissioning this has proven sufficient for our needs.

As mentioned before, insuring we are always able to get the special calibration stream is critical for us to meet the 48 hour turn-around in providing calibration data back to CERN.  In practice we want to achieve 24 hour turn-around.   To improve the resiliency of the system we have configured a protected virtual circuit of 288 Mbits/sec between CERN and our calibration center which can be used as a backup path if there are problems with the primary data distribution.   Future work will try to leverage this circuit for the data returned to CERN via Oracle STREAMS.

To provide the required storage for incoming data the existing Tier-2 dCache system was used to create a new storage area called {\bf AGLT2\_CALIBDISK}.  This area currently provides 30 TB of storage dedicated to the calibration center.

The muon calibration results are sent back to the ATLAS collaboration via replication of the calibration center data back to CERN.   This is achieved by using Oracle to store and manage the calibration data at each calibration site and setting up Oracle STREAMS replication to a common Oracle server at CERN.  This has worked well but it does require that the sites have the expertise to install, configure and manage Oracle as a critical service. The Oracle server at Michigan is hosted an a robust, dedicated Dell PE2950 server with dual E5440 processors, redundant power supplies, bonded gigabit-ethernet network interface cards and 2 TB of RAID10 configured disk (4 TB raw).

The last unique requirement is that the ATLAS muon calibration group is using software developed by Alessandro de Salvo/Rome to manage the calibration workload.  This software (called the ``ATLAS Muon Calibration Data Splitter'') integrates with the local site DDM and job management systems and provides a secure web interface to the calibration work-flow. There are a number of advantages to this including the ability to access the site from any web location, provide a common interface for all three calibration sites and easy management and tracking of the calibration workflow.  Therefore, a new virtual machine dedicated to running the splitter software has been setup at Michigan to provide an easy to manage splitter instance.

\section{Commissioning the Calibration Center}

During the last year we have worked on commissioning our calibration center and we have encountered a number of issues in this process:

\begin{itemize}
\item Integrating the splitter with our local site configuration
\item Oracle maintenance and upgrades in the presence of STREAMS replication
\item Data management issues, primarily related to timely access to data
\item Calibration jobs having slow job initialization due to remote DB access
\end{itemize}

All of these issues were resolved but it is worthwhile to briefly discuss them.

The splitter workflow management software is very useful in tracking the calibration task status and providing a common interface for all three calibration sites.   However there are some intricacies in correctly integrating the splitter configuration with the local site configuration.   The splitter software was developed at Rome using the LCG/EGEE software stack, while AGLT2 uses the Open Science Grid stack. Functionally the stacks are similar and many of the services can interoperate.  However there are a number of small differences that must be accounted for when configuring the splitter to be able to correctly access the needed local site services.   Many of these issues revolve around either grid-security and access or versions of libraries or Python that exist on the site. None of the issues were difficult but they did need to be identified and fixed.

Installing, configuring and maintaining Oracle requires some level of expertise which was not initially present at our site.  We had had some experience  with database technologies in general but needed to acquire more practical experience in the typical tasks associated with running Oracle.   We relied heavily upon advice from the CERN Oracle team as well as utilizing Oracle's MetaLink. However most of our problems resulted from not properly understanding how to patch or upgrade Oracle in the presence of STREAMS replication.   At least two times our efforts to patch our Oracle instance resulted in a corrupted or halted STREAMS replication that required intervention by the CERN Oracle experts to repair.  The take home lesson here has been to better consult with our colleagues about the required steps when undertaking patching or upgrading our Oracle instance.

\begin{figure}[h]
\centering
\includegraphics[width=80mm]{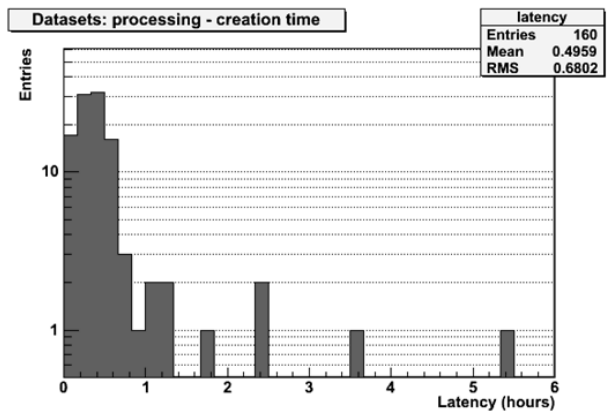}
\caption{This shows the latency (in hours) between data creation at CERN and its arrival at AGLT2.} \label{CosmicLatency}
\end{figure}

There were also a number of minor issues related to the timeliness of data arrival at our calibration center from CERN.  In general we have this working well but there have been occasions during the last year where files were significantly delayed.  Most of these issues were traced to upgrades or changes in the ATLAS DDM system and were quickly resolved.  Tests of the latency of cosmic dataset arrivals are shown in figure~\ref{CosmicLatency}.  Note that even the worst-case time of 5.5 hours should not prevent our center from processing the data within a 24 hour time window.

The last issue we enountered recently (during summer of 2009) was a very slow initialization time for some of our calibration jobs.   Using ATHENA release 14.5.0.1 to do moun reconstruction for 100 events resulted in the processing times shown in table~\ref{proctimes}.

\begin{table}[h]
\begin{center}
\caption{Processing times by site for 100 muon events}
\begin{tabular}{|l|c|c|}
\hline \textbf{Site} & \textbf{Node} & \textbf{Processing Time} 
\\
\hline CERN & pcatum11.cern.ch & 24 minutes \\
\hline BNL & acas002.bnl.gov & 14 minutes \\
\hline AGLT2 & umt3int02.aglt2.org & {\bf 64 minutes} \\
\hline
\end{tabular}
\label{proctimes}
\end{center}
\end{table}

\begin{figure}[h]
\centering
\includegraphics[width=80mm]{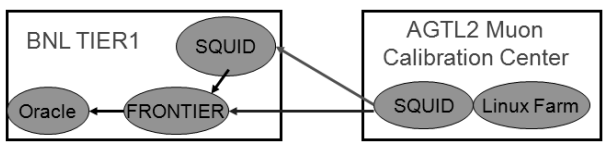}
\caption{This diagram shows the configuration we used to address the slow job initialization when accessing conditions data across the wide-area network.} \label{squid}
\end{figure}

The problem was traced to very slow DB access due to the wide-area network latency from AGLT2 to BNL.  The BNL and CERN runs had ``local'' conditions data available.  We discussed the problem with our colleagues at BNL and decided to try a combination of SQUID\cite{SquidRef} caches and FRONTIER\cite{FrontierRef} to minimize the impact of having to access the database across the network.  Figure~\ref{squid} shows the solution we used.

\begin{figure}[h]
\centering
\includegraphics[width=80mm]{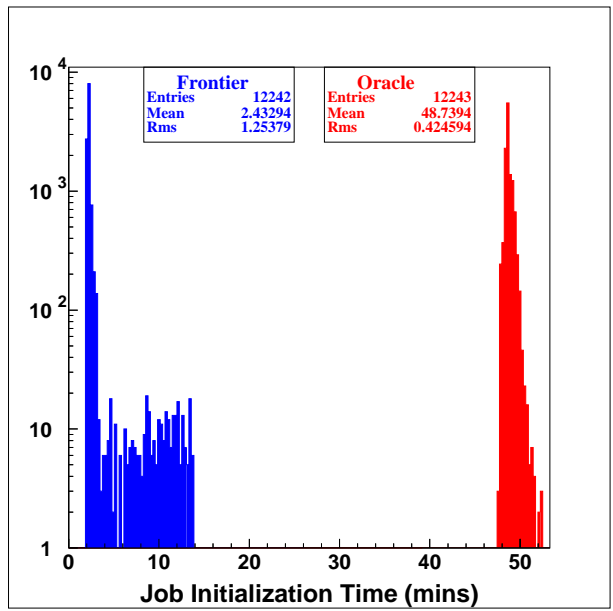}
\caption{The red plot shows the job initialization time distribution using direct Oracle access (on the right) while the blue plot (on the left) shows the corresponding runtime using SQUID/FRONTIER. } \label{frontier}
\end{figure}

The idea is to utilize SQUID's caching ability to minimize the impact of the wide-area network latency.   The results are shown in figure~\ref{frontier} where the mean time for a job decreased from 48.7 minutes to 2.5 minutes.

As part of our testing with our BNL colleagues we tried significantly ramping up the number of jobs utilizing our SQUID/FRONTIER configuration and found that we needed approximately one local SQUID server per 1000 CPUs (jobs).  For AGLT2 we have implemented a set of two SQUID servers and a DNS alias which serves them in a round-robin way. 

\vspace{-10pt}
\section{Future Work and Conclusions}

There are a number of near-term issues we hope to address before the LHC turn-on at the end of the year.   First we need to better integrate the muon splitter job-submission with the ATLAS PANDA system to properly account for and prioritize these jobs.  Currently the splitter directly submits calibration jobs to Condor, bypassing PANDA but this prevents us from properly accounting for these service jobs.

A second task is to work on the secondary data path to ensure robust operation.   We need to update the storage system at CERN to allow it to subscribe to the muon calibration stream and thereby provide a secondary source for this data in the event of problems with the primmary data distribution mechanism.  In addition we would like to take advantage of the protected virtual circuit that exists between CERN and Michigan for the Oracle STREAMS replication. 

Other focus areas will be on continued testing and further integration of the splitter with the local site capabilities.  An important component in this is the addition of more MDT data quality assurance components to help us in the testing and verification of our calibration results.  We should note that the algorithms used for calibration are well developed and tested.

In summary, the AGLT2 computing center is operational and performing very well.  The Michigan ATLAS Muon Calibration center has been successfully deployed and commissioned as a prioritized subset of AGLT2.  While some issues related to our required services and their operation have been found, all of them have been succesfully addressed.   Further testing and improvements are underway to increase the robustness of our centers but the fundamental tools and infrastructure are in place to provide muon calibration results when ATLAS resumes running at the end of 2009.
\vspace{-10pt}
\begin{acknowledgments}
We would like to acknowledge the support of both the Department of Energy and the National Science Foundation in funding the work described here.
\end{acknowledgments}


\end{document}